\documentclass[11pt,a4paper]{article}

\usepackage[bottom=0.5in,top=0.4in]{geometry}
\usepackage{graphicx}
\usepackage{amsmath}
\usepackage{amssymb}
\usepackage{mathrsfs}
\usepackage{microtype}
\usepackage{makeidx}
\usepackage{slashed}
\usepackage[hang,small,bf]{caption}
\usepackage[colorlinks=true,linkcolor=black,citecolor=black,filecolor=black,urlcolor=black,pdftex,
            pdfauthor={Chrysoula Markou},
            pdftitle={Nonlinear},
            pdfsubject={MSc Thesis},
            pdfkeywords={SuGra},
            pdfproducer={Latex with hyperref},unicode={true},
            bookmarksopen={true},bookmarksopenlevel=\maxdimen,bookmarksnumbered={true},
            pdfcreator={pdflatex}]{hyperref}
\usepackage{verbatim}
\usepackage{setspace}
\usepackage{indentfirst}
\usepackage{array}
\usepackage{titlesec}
\usepackage[titletoc, toc, title]{appendix}

\makeindex
            
\numberwithin{equation}{section}
\numberwithin{figure}{section}

\newcommand*\diff{\mathop{}\!\mathrm{d}}

\newcommand{\mi}{\mathrm{i}}

\author{\Large Ignatios Antoniadis$^{a,b,1}$, Chrysoula Markou$^{a,2}$}
\title{\bf The coupling of Non-linear Supersymmetry to Supergravity}

\begin{document}

\begin{titlepage}

\maketitle
\begin{center}
\renewcommand{\thefootnote}{\fnsymbol{footnote}}\vspace{-0.5cm}
${}^{a}$ LPTHE, UMR CNRS 7589, Sorbonne Universit\'es, UPMC Paris 6, 75005 Paris, France\\[0.2cm]
${}^{b}$ Albert Einstein Center for Fundamental Physics, Institute for Theoretical Physics\\ University of Bern, Sidlestrasse 5, 3012 Bern, Switzerland\\[0.5cm]
 \textit{E-mails}:  ${}^{1}$ antoniad@lpthe.jussieu.fr, ${}^{2}$ chrysoula@lpthe.jussieu.fr
\end{center}

\begin{abstract} 

We study the coupling of non-linear supersymmetry to supergravity. The goldstino nilpotent superfield of global supersymmetry coupled to supergravity is described by a geometric action of the chiral curvature superfield $\mathscr{R}$ subject to the constraint $(\mathscr{R}-\lambda)^2=0$ with an appropriate constant $\lambda$. This constraint can be found as the decoupling limit of the scalar partner of the goldstino in a class of $f(\mathscr{R})$ supergravity theories.

\end{abstract}

\providecommand{\keywords}[1]{\textbf{Keywords:} #1}
\thispagestyle{empty}
\end{titlepage}

\section{Introduction}

Studies of non-linear supersymmetric actions have been revived recently due to potential interesting applications in particle physics~\cite{Antoniadis:2010hs} and cosmology~\cite{Antoniadis:2014oya,Ferrara:2014kva,Bergshoeff:2015tra} and their realization in particular string compactifications~\cite{Antoniadis:1999xk,Antoniadis:2004uk}. Indeed, effective actions with non-linear supersymmetry parametrize in a model independent way the effects of supersymmetry breaking at low energies compared to the mass of the sgoldstino (supersymmetric partner of the goldstino) which is in general of the order of the supersymmetry breaking scale at the `hidden' sector. The goldstino on the other hand, although part of the massive gravitino providing its longitudinal degrees of freedom, 
is always in the low-energy spectrum since it becomes massless in the absence of gravity and interacts with a strength fixed by the supersymmetry breaking scale, in contrast to the transverse gravitino components that become free and decouple. In the string theory context, non-linear supersymmetry appears naturally on the world-volume of D-branes realizing the broken supersymmetries of the bulk. It can then even remain an exact symmetry of certain vacua, if for instance the orientifold projection respects it~\cite{Antoniadis:1999xk,Kallosh:2015nia}.

At the global level, a convenient way to write off-shell non-linear supersymmetric actions is by utilizing a nilpotent chiral superfield~\cite{Rocek:1978nb,Lindstrom:1979kq,Casalbuoni:1988xh,komar}. In analogy to ordinary non-linear sigma-models, the constraint eliminates the sgolsdtino component, playing the role of the radial Higgs mode, and replaces it by a goldstino bilinear. In the absence of matter fields, the low-energy action (i.e. without higher order super-derivatives) is completely determined in terms of the goldstino decay constant (or equivalently the supersymmetry breaking scale), and reproduces~\cite{komar, kuzenko} the Volkov-Akulov action~\cite{Volkov:1973ix} on-shell. Indeed, the most general K\"ahler potential is canonical and the superpotenial is linear in the nilpotent goldstino superfield. In the presence of matter, the simple nilpotent constraint may change if some matter fields have superheavy superpartners (of the order of the sgoldstino mass)~\cite{Antoniadis:2011xi}, but it remains valid if all other extra fields belong to ordinary linear supermultiplets~\cite{Antoniadis:2010hs}. 

The coupling to supergravity is straightforward since the constraint does not involve any derivatives~\cite{Antoniadis:2014oya}. The superpotential now admits also a constant piece, allowing for an arbitrary cosmological constant of any sign and space-time to be anti-de Sitter, de Sitter or flat. In flat space, the gravitino mass is given by the usual relation in terms of the supersymmetry breaking scale and the Planck mass. In the unitary gauge, the action is reduced to the ordinary $N=1$ supergravity with a mass term for the gravitino that has absorbed the goldstino. The theory has an alternative geometric formulation in terms of the chiral curvature superfield $\mathscr{R}$ that obeys an appropriate quadratic constraint $(\mathscr{R}-\lambda)^2=0$ with $\lambda$ a constant~\cite{Antoniadis:2014oya,Markou,Dudas:2015eha}.

In this work, we first show the equivalence of the two formulations of non-linear supersymmetry coupled to supergravity by computing explicitly the two actions in components. An alternative way to obtain the constraint is to add it in the action with an independent coupling-coefficient $\rho$ and take the limit $\rho\to\infty$. The resulting $\mathscr{R}^2$ supergravity contains besides the graviton and the gravitino the degrees of freedom of a chiral multiplet that should play the role of the goldstino multiplet. It turns out however, that this theory does not have a minimum in flat space for finite $\rho$, while starting from a de Sitter minimum, the decoupling limit of the sgoldstino, and thus of non-linear supersymmetry, does not exist. We then study a general class of $f(\mathscr{R})$ $N=1$ supergravity theories~\cite{Gates:2009hu, Ketov:2009wc} that satisfy the required limit in flat space; the mass of the extra complex scalar goes to infinity in a Minkowski minimum of the scalar potential and the geometric constraint for the chiral curvature is recovered.

The structure of the paper is the following. In Section 2, we show the equivalence between the two formulations of non-linear supersymmetry coupled to $N=1$ supergravity using first a formal argument with superfields and then by comparing the two effective actions in components. We find two possible values for the constant $\lambda$ entering the geometric constraint: $\lambda=0$ and $\lambda=6W_0$ with $W_0$ a constant superpotential. In Section 3, we recover this constraint in the sgoldstino decoupling limit of a particular class of $f(\mathscr{R})$ supergravity theory that we construct as a Taylor series expansion around a flat space minimum of the corresponding scalar potential.
Finally, in Appendix A we give some details for the derivation of the solution of the geometric constraint, while in Appendix B we show why an $\mathscr{R}^2$ supergravity does not have a stable supersymmetry breaking vacuum that reproduces the constraint in a suitable limit.

\section{Two equivalent Lagrangians} \label{sectt}

In the constrained superfield formalism of non-linear supersymmetry, the goldstino is described by the fermionic component of a chiral superfield $X$, that satisfies the nilpotent constraint $X^2=0$~\cite{Rocek:1978nb,Lindstrom:1979kq,Casalbuoni:1988xh,komar}. The scalar component (sgoldstino) is then eliminated by the constraint and is replaced by a goldstino bilinear. The most general low energy (without super-derivatives) Lagrangian, invariant (upon space-time integration) under global supersymmetry, is then given by
\begin{equation}
\mathscr{L}_{VA} = [X \bar{X}]_D + \left([fX]_F+ \textrm{h.c.}\right),
\label{valag}
\end{equation}
where $f \neq 0$ is a complex parameter. The subscripts $D$ and $F$ denote D and F-term densities, integrated over the full or the chiral superspace, respectively, and correspond to the K\"ahler potential and superpotential of $N=1$ supersymmetry. It can be shown~\cite{komar, kuzenko} that $\mathscr{L}_{VA}$ is equivalent to the Volkov-Akulov Lagrangian \cite{Volkov:1973ix} on-shell. 

The coupling to supergravity in the superconformal context~\cite{super}, \cite{freed}, (\ref{valag}) takes the form
\begin{equation}
\mathscr{L} = -\left[(1 - X \bar{X}) S_0 \bar{S}_0\right]_D + \left( [( fX + W_0 + \frac{1}{2} T X^2 ) S_0^3]_F + \textrm{h.c.}\right),
\label{suoc}
\end{equation}
where we have used the superconformal tensor calculus \cite{cec}, \cite{ferr} with $S_0$ being the superconformal compensator superfield. We have also used a Lagrange multiplier $T$ in order to impose the constraint $X^2 = 0$ explicitly in $\mathscr{L}$, while the factor $\frac{1}{2}$ is put merely for convenience. $W_0$ is a complex constant parameter whose importance will appear shortly. The K\"ahler potential corresponding to (\ref{suoc}) is given by
\begin{equation}
K(X, \bar{X}) = -3 \ln (1 - X \bar{X}) = -3 \left[-X \bar{X} - \frac{(- X \bar{X})^2}{2} + \dots\right] = 3 X \bar{X}.
\label{kahl}
\end{equation}

We would now like to find a geometrical formulation of (\ref{suoc}), that is, to eliminate $X$ and write an equivalent Lagrangian that contains only superfields describing the geometry of spacetime, such as the superspace chiral curvature $\mathscr{R}$~\cite{Antoniadis:2014oya,Markou,Dudas:2015eha}. For that, we observe that the following K\"ahler potential $K'$:
\begin{equation}
K ' = -3 \ln (1 + X +\bar{X}) = -3 \left(X + \bar{X}  - \frac{(X +\bar{X})^2}{2} + \dots\right) = 3 X \bar{X} - 3(X + \bar{X}),
\end{equation}
is related to the K\"ahler potential $K$ via a K\"ahler tranformation of the type
\begin{equation}
\begin{gathered}
K \rightarrow K' = K - 3 (X + \bar{X}) \\
W \rightarrow W' = e^{3X}W.
\end{gathered}
\end{equation}
This tells us that $\mathscr{L}$ is equivalent to $\mathscr{L}'$, where
\begin{equation}
 \mathscr{L}' = -\left[(1 + X  + \bar{X}) S_0 \bar{S}_0\right]_D + \left( [( fX + W_0 + \frac{1}{2} T X^2) e^{3X} S_0^3]_F + \textrm{h.c.}\right).
\end{equation}
Using the constraint $X^2 = 0$, we have 
\begin{align}
 \mathscr{L}' & = -\left[(1 + X  + \bar{X}) S_0 \bar{S}_0\right]_D + \left( [( fX + W_0(1+3X)+ \frac{1}{2} T X^2) S_0^3]_F + \textrm{h.c.}\right)  \nonumber \\
& = - [S_0 \bar{S}_0]_D + \left( [( \lambda X + W_0 - X \frac{\mathscr{R}}{S_0} + \frac{1}{2} T X^2) S_0^3]_F + \textrm{h.c.}\right),
\label{neo}
\end{align}
where we have set $\lambda = f + 3 W_0 $ and we have used the identity \cite{ferr}
\begin{align}
 [ X \cdot \mathscr{R} \cdot S_0^2 ]_F = \left[S_0 \bar{S}_0 (X+ \bar{X})\right]_D + \textrm{total derivatives}.
\label{iwe}
\end{align}

In (\ref{neo}), $X$ enters only in F-terms without derivatives and can be thus integrated out.
Solving the equation of motion for $X$, we have 
\begin{equation}
 \lambda - \frac{\mathscr{R}}{S_0} + TX = 0\quad \Rightarrow\quad X = \frac{\frac{\mathscr{R}}{S_0} - \lambda}{T}
\end{equation}
and substituting back into (\ref{neo}), we get 
\begin{align}
  \mathscr{L}' & = - [S_0 \bar{S}_0]_D + \left( [( - \frac{1}{2T} (\frac{\mathscr{R}}{S_0} - \lambda  )^2 + W_0 ) S_0^3]_F + \textrm{h.c.}\right) \nonumber \\
& = \left[(- \frac{1}{2}\frac{\mathscr{R}}{S_0} + W_0 - \frac{1}{2T} (\frac{\mathscr{R}}{S_0} - \lambda  )^2 ) S_0^3  \right]_F + \textrm{h.c.} \, ,
\label{dua}
\end{align}
where we have used again the identity (\ref{iwe}). We can now view $\frac{1}{T}$ as a Lagrange multiplier that imposes the constraint
\begin{equation}
 (\frac{\mathscr{R}}{S_0} - \lambda  )^2 = 0 \, .
\label{con}
\end{equation}
Consequently, we have established an equivalence between the constrained Lagrangians (\ref{suoc}) and (\ref{dua}); they both describe the coupling of non-linear supersymmetry to supergravity, with $\mathscr{L}'$ providing its geometric formulation with the use of a constraint imposed on $\mathscr{R}$ instead of $X$. This constraint was proposed in~\cite{Antoniadis:2014oya} for $\lambda=0$. In what follows we will confirm the equivalence by writing these Lagrangians in terms of component fields.

\subsection{Constraining a chiral superfield $X$}

In the following we use the method and conventions of \cite{wesbag} except from a factor of $1/6$ which we omit in the expression of $\mathscr{R}$ but introduce at the Lagrangian level. We also set the gravitational coupling $\kappa^2 = 8 \pi G_N$ (given here in natural units) to be equal to one, in accordance with the usual convention. After gauge-fixing the superconformal symmetry by using the convenient gauge $S_0=1$, the Lagrangian (\ref{suoc}) can be written as follows:
\begin{equation} 
\begin{gathered}
\mathscr{L} = \int \diff^2 \Theta 2 \mathscr{E} \left\{  \frac{3}{8} (\bar{\mathscr{D}} \bar{\mathscr{D}} - \frac{8}{6} \mathscr{R}) e^{-K/3} + W \right\} + \textrm{h.c.} \\
{\rm with}\ W(X) = fX + W_0\ {\rm and}\ X^2 =0 \, ,
\end{gathered}
\label{xx}
\end{equation}
where $\mathscr{D}$ is the super-covariant derivative and $\mathscr{E}$ the chiral superfield density that is constructed from the vielbein $e_a^m$:
\begin{equation}
\mathscr{E} = \frac{1}{2}e \left\{ 1 + i\Theta \sigma^a \bar{\psi}_a - \Theta \Theta [\bar{M} + \bar{\psi}_a \bar{\sigma}^{ab} \bar{\psi}_b] \right\}\, .
\end{equation}
Here $\psi_a$ is the gravitino, $\Theta$ the fermionic coordinates of the curved  superspace and $\sigma^a=(-1,\vec\sigma)$, $\sigma^{ab \beta}_{\alpha}=\frac{1}{4}(\sigma^a_{\alpha \dot{\alpha}} \bar{\sigma}^{b \dot{\alpha} \beta}  - \sigma^b_{\alpha \dot{\alpha}} \bar{\sigma}^{a \dot{\alpha} \beta}    ) $  with $\vec\sigma$ the Pauli matrices. Note that the Lagrange multiplier $T$ in (\ref{suoc}) has been used to impose the constraint $X^2 =0$, which can be solved, fixing the scalar component (sgoldstino) in terms of the goldstino $G$ and the auxiliary field $F$ of X \cite{komar}. 

We now substitute $X$, $\mathscr{E}$ and $\mathscr{R}$ with their respective expressions in component fields: 
\begin{align}
\begin{gathered}
X  = \frac{G^2}{2F} + \sqrt 2 \Theta G + (\Theta \Theta)F \equiv A + \sqrt 2 \Theta G + (\Theta \Theta)F \\
\mathscr{R} \equiv - M - \Theta B -  (\Theta \Theta) C \\
\Xi \equiv (\bar{\mathscr{D}} \bar{\mathscr{D}} - \frac{8}{6} \mathscr{R}) \bar{X}  \equiv - 4 \bar{F} + \frac{4}{3} M \bar{A} + \Theta D + (\Theta \Theta) E \, .
\label{compo}
\end{gathered}
\end{align}
The exact components of $\mathscr{R}$ and $\Xi$ are computed in~\cite{wesbag} (our convention for $\mathscr{R}$ differs by $1/6$ with respect to \cite{wesbag}). $M$ and $b_a$ are the auxiliary fields of the $N=1$ supergravity multiplet in the old-minimal formulation. Then
\begin{align}
-\frac{3}{4}\left[\mathscr{E} ( \bar{\mathscr{D}} \bar{\mathscr{D}} - \frac{8}{6} \mathscr{R}) \bar{X} X   \right]_F = - \frac{3}{8} [2 \mathscr{E} \Xi X]_F = - \frac{3}{8} e (EA - 4 F \bar{F} + \frac{4}{3}  M F \bar{A}  - \frac{\sqrt 2}{2} (DG)) \nonumber \\
+ \frac{3}{16} \mi e (y \sigma^a \bar{\psi}_a) + \frac{3}{8}e[\bar{M} + \bar{\psi}_a \bar{\sigma}^{ab} \bar{\psi}_b] [ -4 A \bar{F} + \frac{4}{3} M A \bar{A} ] \, ,
\end{align}
where
\begin{equation}
\begin{gathered}
y = \sqrt 2 G (-4 \bar{F} + \frac{4}{3}M \bar{A}) + DA \, .
\end{gathered}
\end{equation}
This expression is simplified significantly if we choose to use the unitary gauge, setting $G = 0$ and thus $A = y = 0$:
\begin{equation}
-\frac{3}{4}\left[\mathscr{E} ( \bar{\mathscr{D}} \bar{\mathscr{D}} - \frac{8}{6} \mathscr{R}) \bar{X} X   \right]_F = \frac{3}{2} e F \bar{F} \, .
\end{equation}

Moreover, also in the unitary gauge, one can compute 
\begin{equation}
\left[ 2 \mathscr{E} (fX + W_0)\right]_F = ef F - e \left[\bar{M}+ \bar{\psi}_a \bar{\sigma}^{ab} \bar{\psi}_b\right] W_0\, .
\end{equation}
Now, using the property
\begin{equation}
(\bar{\psi}_a \bar{\sigma}^{ab} \bar{\psi}_b)^{\dagger} = \frac{1}{4} \left[\bar{\psi}_a (\bar{\sigma}^a \sigma^b - \bar{\sigma}^b \sigma^a )  \bar{\psi}_b\right]^{\dagger} = \frac{1}{4} \left[ \psi_b (\sigma^b \bar{\sigma}^a - \sigma^a \bar{\sigma}^b) \psi_a \right] = \psi_a \sigma^{ab} \psi_b \, ,
\end{equation}
the Lagrangian (\ref{xx}) in terms of component fields becomes
\begin{align}
\mathscr{L} =   -\frac{1}{2}eR - \frac{1}{3} eM \bar{M} + \frac{1}{3}eb^a b_a +\frac{1}{2}e \epsilon^{abcd} ( \bar{\psi}_a \bar{\sigma}_b \widetilde{\mathscr{D}}_c \psi_d -\psi_a \sigma_b \widetilde{\mathscr{D}}_c {\bar{\psi}}_d   ) \nonumber \\
 + efF - e W_0  [ \bar{M} + \bar{\psi}_a \bar{\sigma}^{ab} \bar{\psi}_b  ] +  e \bar{f} \bar{F}- e \bar{W_0} [ M + \psi_a \sigma^{ab} \psi_b  ]   + 3e F \bar{F}\, ,
\label{tel}
\end{align}
where $R$ is the Ricci scalar.
The equations of motion for the auxiliary fields $b^a$, $M$, $F$ are then
\begin{equation}
\begin{gathered}
b^a = 0 \\
M = -3 W_0,\bar{M} = -3 \bar{W_0} \\
F = - \frac{\bar{f}}{3}, \bar{F} = - \frac{f}{3}. 
\end{gathered}
\end{equation}
Substituting back into (\ref{tel}) we get
\begin{eqnarray}
\mathscr{L} &=& -\frac{1}{2}e R+\frac{1}{2}e \epsilon^{abcd} ( \bar{\psi}_a \bar{\sigma}_b \widetilde{\mathscr{D}}_c \psi_d -\psi_a \sigma_b \widetilde{\mathscr{D}}_c {\bar{\psi}}_d   )  \nonumber \\
&&- e W_0  \bar{\psi}_a \bar{\sigma}^{ab} \bar{\psi}_b - e \bar{W_0} \psi_a \sigma^{ab} \psi_b +  3e |W_0|^2 - \frac{1}{3}e|f|^2.
\label{sR}
\end{eqnarray}
In this form, it is obvious that the Lagrangian reduces to the usual $N=1$ supergravity, together with a gravitino mass term: 
\begin{equation}
\boxed{m_{3/2} = |W_0|}.
\end{equation}

Imposing that the cosmological constant (i.e. the vacuum expectation value of the scalar potential) vanishes, one finds
\begin{equation}
 3 |W_0|^2 - \frac{1}{3}|f|^2 = 0 \Rightarrow \boxed{ |f|^2 = 9|W_0|^2 }.
\label{sam}
\end{equation}
This means that $W_0 \neq 0$, which justifies the use of the constant piece $W_0$ in the superpotential 
in $\mathscr{L}$. Then, the final form of $\mathscr{L}$ is 
\begin{align}
\mathscr{L} = -\frac{1}{2}e R+\frac{1}{2}e \epsilon^{abcd} ( \bar{\psi}_a \bar{\sigma}_b \widetilde{\mathscr{D}}_c \psi_d -\psi_a \sigma_b \widetilde{\mathscr{D}}_c {\bar{\psi}}_d   )  
- e W_0  \bar{\psi}_a \bar{\sigma}^{ab} \bar{\psi}_b - e \bar{W_0} \psi_a \sigma^{ab} \psi_b \, .
\label{fin}
\end{align}
It is important to notice that the use of the constrained superfield $X$ is what has generated the gravitino mass term: the final form of the Lagrangian in flat space is just the pure $N=1$ supergravity, but with a massive gravitino. The use of the unitary gauge $G  = 0$ results in the gravitino absorbing the goldstino and becoming massive, in analogy with the well-known Brout-Englert-Higgs mechanism.

\subsection{Constraining the superspace curvature superfield $\mathscr{R}$}

After gauge-fixing the superconformal symmetry by imposing $S_0=1$, the Lagrangian (\ref{dua}) can be written as follows:
\begin{equation} 
\begin{gathered}
 \mathscr{L}' = - \int d^2 \Theta \mathscr{E} (\mathscr{R} - 2 W_0) + \textrm{h.c.}, \\
 (\mathscr{R} - \lambda)^2 = 0.
\end{gathered}
\label{yy}
\end{equation}
$\mathscr{L}'$ then yields
\begin{align}
\mathscr{L}' =  -\frac{1}{2}eR - \frac{1}{3} eM \bar{M} + \frac{1}{3}eb^a b_a +\frac{1}{2}e \epsilon^{abcd} ( \bar{\psi}_a \bar{\sigma}_b \widetilde{\mathscr{D}}_c \psi_d -\psi_a \sigma_b \widetilde{\mathscr{D}}_c {\bar{\psi}}_d   ) \nonumber \\
-e W_0 [\bar{M} + \bar{\psi}_a \bar{\sigma}^{ab} \bar{\psi}_b] - e \bar{W}_0 [M + \psi_a \sigma^{ab} \psi_b].
\label{duall}
\end{align}
Now let us solve the constraint which is the second of the equations (\ref{yy}). For that, we substitute the second of the equations (\ref{compo}) into the constraint and find the set of the following equations:
\begin{equation}
 \begin{gathered}
  (M + \lambda)^2 = 0 \\
   (M + \lambda)B_{\alpha} = 0 \\
    4 (M+ \lambda) C = (BB),
\label{werd}
 \end{gathered}
\end{equation}
where
\begin{equation}
\begin{gathered}
B_{\alpha} = \sigma^{a}_{\alpha \dot{\alpha}} \bar{\sigma}^{b \dot{\alpha} \beta} \psi_{ab \beta} - \mi  \sigma^a_{\alpha \dot{\alpha}} \bar{\psi}_{a}^{\dot{\alpha}} M + \mi \psi_{a \alpha} b^a 
\quad{\rm with}\quad \psi_{ab} \equiv \widetilde{\mathscr{D}}_a\psi_b - \widetilde{\mathscr{D}}_b\psi_a\\
C = - \frac{1}{2} R + \mathcal{O} \{M, b_a, \psi_a \} \neq 0\, .
\label{BandC}
\end{gathered}
\end{equation}
Equations (\ref{werd}) yield:
\begin{equation}
M =  - \lambda 
\quad {\rm and}\quad b^a = 0\, .
\label{conn}
\end{equation}
Indeed, $B$ in this case depends only on the gamma-trace or the divergence of the gravitino, $\bar{\sigma}^a\psi_a$ and $\widetilde{\mathscr{D}}^a\psi_a$ (using the Clifford algebra property of sigma-matrices $(\sigma^a\bar\sigma^b+\sigma^b\bar\sigma^a)_{\alpha}^{\beta}  = - 2\eta^{ab} \delta_{\alpha}^{\beta}$) 
, that can be put to zero by an appropriate gauge choice. Alternatively, one can show that $B$ vanishes on-shell (see Appendix A).

Using (\ref{conn}), eq.~(\ref{duall}) becomes: 
\begin{align}
\mathscr{L}' =  -\frac{1}{2}eR  - \frac{1}{3} e|\lambda|^2  +\frac{1}{2}e \epsilon^{abcd} ( \bar{\psi}_a \bar{\sigma}_b \widetilde{\mathscr{D}}_c \psi_d -\psi_a \sigma_b \widetilde{\mathscr{D}}_c {\bar{\psi}}_d   ) \nonumber \\
+ e W_0 \bar{\lambda} + e \bar{W}_0 \lambda -e W_0 \bar{\psi}_a \bar{\sigma}^{ab} \bar{\psi}_b - e \bar{W}_0 \psi_a \sigma^{ab} \psi_b \, .
\label{duallf}
\end{align}
Substituting now $\lambda=f + 3W_0$, one finds that the cosmological constant is given by $3e|W_0|^2-\frac{1}{3}e|f|^2$ and the Lagrangian (\ref{duallf}) is identical to (\ref{sR}). Note that the vanishing of the cosmological constant
\begin{equation}
 - \frac{1}{3} e|\lambda|^2 + e W_0 \bar{\lambda} + e \bar{W}_0 \lambda = 0
\label{same}
\end{equation}
gives two possible solutions for $\lambda$: 
\begin{equation}
\boxed{\lambda=6 W_0}\quad {\rm and}\quad \boxed{\lambda=0}\, , 
\end{equation}
corresponding to $f=\pm 3W_0$ that solve the condition (\ref{sam}).

\section{Without imposing direct constraints} \label{tria}

In this section, we would like to start with a regular $\mathscr{R}^2$ supergravity and recover the constraint in an appropriate limit where the additional (complex) scalar arising from $\mathscr{R}^2$ becomes superheavy and decouples from the low energy spectrum. Indeed, by analogy with ordinary General Relativity in the presence of an $R^2$-term (with $R$ the scalar curvature), an $\mathscr{R}^2$ supergravity can be re-written as an ordinary Einstein $N=1$ supergravity coupled to an extra chiral multiplet.\footnote{Note that $\mathscr{R}^2$ supergravity is not the supersymmetrization of $R^2$ gravity which is described by a D-term $\mathscr{R}\bar{\mathscr{R}}$, bringing two chiral multiplets to be linearized~\cite{cec,ferr}.}
Let us then consider the Lagrangian
\begin{equation}
 \bar{\mathscr{L}} = \left[ \left(- \frac{1}{2} \frac{\mathscr{R}}{S_0} + W_0 + \frac{1}{2} \rho (\frac{\mathscr{R}}{S_0} - \lambda)^2\right) S_0^3 \right]_F + \textrm{h.c.} \, ,
\label{inn}
\end{equation}
where $\rho$ is a real parameter. In the limit $|\rho| \rightarrow \infty$, one would naively expect to recover the constraint $ (\mathscr{R} - \lambda)^2 \rightarrow 0 $, 
and thus (\ref{inn}) should be reduced to (\ref{dua}). In principle, one could linearize 
(\ref{inn}) with the use of a chiral superfield $S$ and then demonstrate that in the limit  $|\rho| \rightarrow \infty $, $\mathscr{L}$, $\mathscr{L}'$ and $ \bar{\mathscr{L}}$ are all equivalent. If this were true, 
one would expect that $S$ corresponds to the goldstino superfield and that supersymmetry is non-linearly realized (in the limit $|\rho| \rightarrow \infty$), as is the case for the chiral nilpotent superfield $X$. In other words, the mass of the scalar component of $S$ would approach 
infinity as $|\rho| \rightarrow \infty$ and would, therefore, decouple from the spectrum. However, upon computing the scalar potential and the scalar mass matrix corresponding to (\ref{inn}), we found that this is not the case.
This means that the parameter space $(\lambda, W_0, \rho)$ does not allow for a supersymmetry breaking minimum that realizes the sgoldstino decoupling and the equivalence between $\bar{\mathscr{L}}$ with $\mathscr{L}$ and $\mathscr{L}'$. The detailed analysis can be found in the Appendix B.

To solve this problem, we start with a more general class of $f(\mathscr{R})$ supergravity actions. More precisely, we modify $ \bar{\mathscr{L}}$ with the addition of a suitable term that is supressed by $\rho$ in the limit $|\rho|\rightarrow\infty$:\footnote{In principle, we may replace $1/\rho$ by $1/\hat\rho(\rho)$ with $|\hat\rho(\rho)|\to\infty$ when $\rho\to\infty$. One can show however that our results do not change and thus we make the simple choice $\hat\rho=\rho$.}
\begin{equation}
 \mathscr{L}'' = \left[ \left(- \frac{1}{2} \frac{\mathscr{R}}{S_0} + W_0 + \frac{1}{2} \rho (\frac{\mathscr{R}}{S_0} - \lambda)^2   + \frac{1}{\rho} \left(S \frac{\mathscr{R}}{S_0} - F(S) \right) \right) S_0^3 \right]_F + \textrm{h.c.} \, ,
\label{ini}
\end{equation}
where $S$ is a chiral superfield coupled to gravity and $F(S)$ is a holomorphic function of the superfield $S$. This extra term has already been studied in the literature and is known as $f(\mathscr{R})$ supergravity \cite{Gates:2009hu,Ketov:2009wc}. Indeed, $S$ can be integrated out by its equation of motion at finite $\rho$:
\begin{equation}
\mathscr{R} = F' S_0,
\label{solveR}
\end{equation}
where $F' = \frac{\partial F}{\partial S}$. This equation can be in principle solved to give $S$ as a function of $\mathscr{R}$ and replacing it back in (\ref{ini}) one finds an $f(\mathscr{R})$ theory. 

We will now study the physical implications of $\mathscr{L}''$ in the limit $\rho \rightarrow \infty$ so as to confirm the equivalence between $\mathscr{L}$, $\mathscr{L}'$ and $\mathscr{L}''$ (without loss of generality, we take $\rho$ positive).
We first use eq. (\ref{solveR}) to replace $\mathscr{R}$ in terms of $S$ in the third term of (\ref{ini}), instead of doing the reverse as described above. Using then the identity (\ref{iwe}), we get 
\begin{equation}
 \mathscr{L}'' = - \left[(1 - \frac{1}{\rho} (S + \bar{S})) S_0 \bar{S}_0 \right]_D + \left\{ [ (W_0 + \frac{1}{2} \rho (F' - \lambda)^2   - \frac{1}{\rho}  F ) S_0^3 ]_F + \textrm{h.c.} \right\}.
\end{equation}
We now fix the gauge according to $S_0 = 1$ and set $\phi$ to be the lowest component of $S$. Then the K\"ahler potential and the superpotential corresponding to $\mathscr{L}''$ are given by
(we use the same symbols $K$ and $W$ as in section (\ref{sectt}) as there is no confusion)
\begin{equation}
\begin{gathered}
 K = -3 \ln \left(1 - \frac{1}{\rho} (\phi + \bar{\phi})\right) \\
  W = W_0 + \frac{1}{2} \rho (F' - \lambda)^2   - \frac{1}{\rho}  F,
\end{gathered}
\end{equation}
where now $F' = \frac{\partial F}{\partial \phi}$.

It follows that
\begin{equation}
 \textrm{exp}(K) = \frac{\rho^3}{(\rho-\phi-\bar{\phi})^3}
\end{equation}
and
\begin{equation}
 g_{\phi \bar{\phi}} = \frac{\partial}{\partial \phi} \frac{\partial}{\partial \bar{\phi}} K = \frac{3}{(\rho- \phi-\bar{\phi})^2}\quad,\quad g^{\phi \bar{\phi}} =   \frac{(\rho- \phi-\bar{\phi})^2}{3}.
\end{equation}
Also
\begin{align}
 D_{\phi}W & = \partial_{\phi} W + K_{\phi} W = \rho F'' (F' - \lambda) - \frac{1}{\rho} F' + \frac{3}{\rho- \phi-\bar{\phi}} \left(W_0 + \frac{1}{2} \rho (F' - \lambda)^2   - \frac{1}{\rho} F\right).
\end{align}
Putting everything together, we get that the scalar potential $V$ is given by:
\begin{align}
V = \exp(K) \left[ g^{\phi \bar{\phi}}  (D_{\phi} W) (\bar{D}_{\bar{\phi}} \bar{W})  -3 \bar{W} W \right] = \frac{\rho^2}{3(\rho- \phi-\bar{\phi})^2} \tilde{V},
\end{align}
where
\begin{align}\label{potential}
\tilde{V}  = \rho^4 |F''(F' - \lambda)|^2 + \rho^3 \left[ -(\phi + \bar{\phi}) |F''(F' - \lambda)|^2   + \frac{3}{2} |F' - \lambda|^2 \left(F''(\bar{F}' - \bar{\lambda}) + \textrm{h.c.}\right)  \right]  \nonumber \\
+ \rho^2 \left[- \bar{F}' F'' (F' - \lambda) + 3 \bar{W}_0  F''\left(F' - \lambda) + \textrm{h.c.}\right)  \right] + \rho \left[(\phi + \bar{\phi})\bar{F}' F'' (F' - \lambda) - 3 \bar{F}F'' (F' - \lambda)\right. \nonumber \\
-\left. \frac{3}{2} F' (\bar{F}' - \bar{\lambda})^2 +     \textrm{h.c.}   \right] + \rho^0 \left[|F'|^2 - 3 F' \bar{W}_0 -3 \bar{F}' W_0  \right] + \rho^{-1} \left[- (\phi + \bar{\phi} )|F'|^2 + 3 F' \bar{F} + 3\bar{F}' F  \right].     \nonumber \\
\end{align}

For $\rho \rightarrow \infty$, the leading behaviour of $V$ is given by
\begin{equation}
V = \frac{\rho^4}{3} |F''(F' - \lambda)|^2.
\end{equation}
It is positive definite with a minimum at zero when $F'=\lambda$ or $F''=0$. In the following, we will analyze the case $F'=\lambda$; its curvature defines the (canonically normalized) scalar mass given by 
\begin{equation}
m_\phi=\frac{\rho^3}{3}(F'')^2
\label{mass}
\end{equation}
which goes to infinity at large $\rho$ and $\phi$ decouples. At the minimum $F'=\lambda$, the potential at large $\rho$ becomes constant, proportional to $|\lambda|^2 - 3 \lambda \bar{W}_0 - 3 \bar{\lambda} W_0$. This term vanishes precisely if equation (\ref{same}), or equivalently (\ref{sam}), holds. We conclude that in the model (\ref{ini}) the cosmological constant can be tuned to zero (in the limit $\rho \rightarrow \infty$) by using the same condition as for the model (\ref{dua}). As shown in Section 2.2, this is the case for two possible values of $\lambda$:
\begin{equation}
\boxed{\lambda = 6 W_0}\quad\textrm{ or }\quad \boxed{\lambda = 0}\, .
\label{what}
\end{equation}

Now let us investigate the minimum of the potential at finite but large $\rho$. We shall construct the solution as a power series in $1/\rho$ around the asymptotic field value of the minimum $\phi=\phi_0$ that solves $F'=\lambda$. A simple inspection of the potential (\ref{potential}) shows that it is sufficient to consider only even powers in $1/\rho$:
\begin{equation}\label{phiexpansion}
 \begin{gathered}
\phi = \phi_0 + \frac{\phi_1}{\rho^2} \\
F'(\phi) = F'(\phi_0) + (\phi - \phi_0)F''(\phi_0) + \frac{1}{2}(\phi - \phi_0)^2 F'''(\phi_0) + \dots
 \end{gathered}
\end{equation}
or equivalently,
\begin{equation}\label{Fexpansion}
 F'(\phi) = \lambda + \frac{c}{\rho^2} + \frac{d}{\rho^4} + ...
\end{equation}
where
\begin{equation}
 c = \phi_1 F''_0\quad,\quad d = \frac{1}{2} \phi_1^2 F'''_0\, .
\end{equation}
We then compute the derivative of $\tilde{V}$ with respect to $\phi$ and keep only the terms that do not vanish in the limit $\rho\rightarrow\infty$:
\begin{align}
\tilde{V}_{\phi} = \frac{\partial \tilde{V}}{\partial \phi} =  \rho^4 (\bar{F}'' F''' |F'-\lambda|^2 + |F''|^2 F'' (\bar{F}' - \bar{\lambda})  ) - \rho^3 (\phi + \bar{\phi}) |F''|^2 F'' (\bar{F}' - \bar{\lambda}) \nonumber \\
+ \rho^2 [F''^2 (3 \bar{W}_0  - \bar{F}' ) - |F''|^2 (\bar{F}' - \bar{\lambda})  + F''' (F' - \lambda)(3 \bar{W}_0 - \bar{F}') ] \nonumber \\
+ \rho F''^2 [\bar{F}' (\phi + \bar{\phi})   - 3 \bar{F}] + \rho^0 F'' [\bar{F}' - 3 \bar{W}_0].
\end{align}
This expression vanishes if every coefficient at each order vanishes.

We now substitute the expansion (\ref{phiexpansion}), (\ref{Fexpansion}) into $\tilde{V}_{\phi}$ (ignoring orders that vanish as $\rho^{-2}$ and higher) and impose each coefficient to be set to zero so as to have an extremum. Assuming for simplicity that $W_0, \lambda, \phi_0, c, d$ are real, we find the following constraints on the function $F$:
\begin{equation}
 \begin{gathered}
  c F_0'' = \lambda -  3 W_0 \\
  F_0 =  2 W_0 \phi_0  \\
  c^2 F_0''' = \frac{2}{3} (\lambda - 3 W_0) \, , 
 \end{gathered}
\end{equation}
which yield
\begin{eqnarray}
F(\phi) &=& 2 \phi_0 W_0 + \lambda (\phi - \phi_0) + \frac{\lambda - 3W_0}{2c} (\phi - \phi_0)^2 + \frac{1}{3!} \frac{2(\lambda -3W_0)}{3c^2} (\phi - \phi_0)^3 + \dots \nonumber\\ 
&=& 2 \phi_0 W_0 + \lambda (\phi - \phi_0) \pm \frac{3W_0}{2c} (\phi - \phi_0)^2 \pm \frac{1}{3!} \frac{2W_0}{c^2} (\phi - \phi_0)^3 + \dots\, ,
\end{eqnarray}
where in the second line above, we used the two possible values of $\lambda$ (\ref{what}), $\lambda=6W_0$ for the $+$ sign and $\lambda=0$ for the $-$ sign, for which the potential vanishes at the minimum.

At the minimum, the F-auxiliary term of $S$, ${\cal F}_\phi$, is given by:
\begin{eqnarray}
\langle |{\cal F}^\phi | \rangle  
=\langle \left| e^{K/2}
\sqrt{g^{\phi \bar{\phi}}} \bar{D}_{\bar{\phi}}\bar{W} \right| \rangle & \overset{\rho \rightarrow \infty}{\longrightarrow}& \frac{\rho^2}{\sqrt{3}} \langle\left| F'' (F'- \lambda) \right|\rangle + \textrm{(subleading terms)} \nonumber\\ 
&=&  \frac{1}{\sqrt{3}} \langle\left| F_0'' c \right| \rangle + {\cal O}(1/\rho^2) =  \frac{1}{\sqrt{3}}  |\lambda -  3 W_0| \\
&=& \sqrt{3} |W_0|  \neq 0 \nonumber \, ,
\end{eqnarray}
where in the third line we used $\lambda=0$ or $\lambda=6W_0$. We conclude that supersymmetry is spontaneously broken in this limit along the direction of $\phi$, which can be identified with the scalar superpartner of the goldstino that becomes superheavy and decouples. 
The supersymmetry breaking scale remains finite and is given by $f= 3|W_0|$. Therefore, we identify the fermionic component of $S$ with the goldstino and $\phi$ with its superpartner, the sgoldstino. According to (\ref{mass}), the latter decouples from the spectrum in the limit $\rho \rightarrow \infty$, which is equivalent to imposing the nilpotent constraint for the goldstino superfield $X^2=0$ on $\mathscr{L}$. Finally, the gravitino mass is given by
\begin{equation}
\boxed{m_{3/2} = \langle |W|e^{K/2} \rangle \rightarrow |W_0|} \textrm{ as } \rho \rightarrow \infty,
\end{equation}
which completes the proof of equivalence between $\mathscr{L}$, $\mathscr{L}'$ and $\mathscr{L}''$.

\section*{Acknowledgements}
I.A. would like to thank E. Dudas for enlightening discussions. C.M. would like to thank the laboratory LPTHE of the UPMC for partial financial support as well as for the warm hospitality in regard to the Master's Thesis. C.M. is also grateful to the Greek State Scholarships Foundation (IKY) for partially funding her Master's Studies through the programme ``IKY Scholarships''.

\bigskip

\appendix
\section{Appendix A}

Here, we derive the equation of motion for the gravitino from (\ref{duallf}):
\begin{equation}
\frac{1}{2} \epsilon^{abcd}  \sigma_b \widetilde{\mathscr{D}}_c {\bar{\psi}}_d  = - \bar{W}_0 \sigma^{ab} \psi_b \, .
\label{contr}
\end{equation}
Contracting (\ref{contr}) with $\widetilde{\mathscr{D}}_a$, we obtain the following equation:
\begin{equation}
\sigma^{ab} \widetilde{\mathscr{D}}_a  \psi_b  = 0 \, .
\label{usef}
\end{equation}
Moreover, contracting the hermitian conjugate 
\begin{equation}
\frac{1}{2} \epsilon^{abcd}  \bar{\sigma}_b \widetilde{\mathscr{D}}_c {\psi}_d  = W_0 \bar{\sigma}^{ab} \bar{\psi}_b \, .
\label{contrt}
\end{equation}
of (\ref{contr}) with $\sigma_a$, we have that
\begin{equation}
 \epsilon^{abcd}  \sigma_a \bar{\sigma}_b \widetilde{\mathscr{D}}_c {\psi}_d  \sim  \epsilon^{abcd}  \sigma_{ab}  \widetilde{\mathscr{D}}_c {\psi}_d \sim \sigma^{cd}  \widetilde{\mathscr{D}}_c {\psi}_d  = 0 \, ,
\end{equation}
where we have used (\ref{usef}) and
\begin{equation}
 \epsilon^{abcd}  \sigma_{ab} = -2 \mi \sigma^{cd}  \, .
\end{equation}
Consequently,
\begin{equation}
\sigma_a  \bar{\sigma}^{ab} \bar{\psi}_b = 0 \Rightarrow \sigma^a \bar{\psi}_a = 0 \, ,
\label{gaug}
\end{equation}
where we have used the identity
\begin{equation}
\sigma^a \bar{\sigma}^b \sigma^c - \sigma^c \bar{\sigma}^b \sigma^a = 2 \mi  \epsilon^{abcd} \sigma_d \, .
\end{equation}

Now let us consider $B_{\alpha}$ of eq.~(\ref{BandC}). Its last term $ \mi \psi_a b^a $ vanishes due to the equation of motion for $b^a$, while its second term vanishes due to equation (\ref{gaug}).  $B_{\alpha}$'s first term is:
\begin{equation}
\sigma^{a} \bar{\sigma}^b \psi_{ab} =  \sigma^{a} \bar{\sigma}^b  (\widetilde{\mathscr{D}}_a \psi_b - \widetilde{\mathscr{D}}_b \psi_a ) = ( \sigma^{a} \bar{\sigma}^b -  \sigma^{b} \bar{\sigma}^a) \widetilde{\mathscr{D}}_a \psi_b = 4 \sigma^{ab} \widetilde{\mathscr{D}}_a \psi_b = 0 \, ,
\end{equation}
where we have used the definition
\begin{equation}
\sigma^{ab} \equiv \frac{1}{4} ( \sigma^{a} \bar{\sigma}^b -  \sigma^{b} \bar{\sigma}^a)
\end{equation}
and the relation (\ref{usef}). Consequently $B_{\alpha} = 0$ on-shell, which justifies the solution $M=-\lambda$ and $b^a=0$ we chose in Section 2.2.

\section{Appendix B}

We will now demonstrate why the Lagrangian
\begin{equation}
 \bar{\mathscr{L}} = \left[ \left(- \frac{1}{2} \frac{\mathscr{R}}{S_0} + W_0 + \frac{1}{2} \rho (\frac{\mathscr{R}}{S_0} - \lambda)^2\right) S_0^3 \right]_F + \textrm{h.c.}
\label{baro}
\end{equation}
does not reproduce (\ref{dua}) with the constraint (\ref{con}) in the limit $\rho\to\infty$. We first set
\begin{equation}
\begin{gathered}
a= W_0 + \frac{1}{2} \rho \lambda^2 \\
b= 1 + 2 \rho  \lambda \, ,
\end{gathered}
\end{equation}
assuming again reality of all parameters for simplicity. We then introduce a chiral superfield 
$$
S=A + \sqrt 2 \Theta \chi+ (\Theta \Theta) F
$$ 
($A$ and $F$ are not the same as in the previous sections) such that
\begin{align}
\bar{\mathscr{L}} & = \left[ \left( a - \frac{1}{2} b \frac{\mathscr{R}}{S_0}  + \frac{1}{2} \rho \frac{\mathscr{R}^2}{S_0^2} \right) S_0^3\right]_F +  \textrm{h.c.} \nonumber \\
& = \left[ \left(a - \frac{1}{2} b \frac{\mathscr{R}}{S_0} + \frac{S}{S_0} \frac{\mathscr{R}}{S_0} - \frac{1}{2 \rho} \frac{S^2}{S_0^2} \right) S_0^3 \right]_F +  \textrm{h.c.}
 \label{lin}
\end{align}
It follows that $b>0$ in order to have canonical gravity for a metric tensor with signature $(- +++)$. 
 It is obvious from (\ref{lin}) that we have linearized our initial theory (\ref{baro}), which now describes the coupling of supergravity to a chiral superfield $S$ that satisfies the equation of motion
\begin{equation}
 S =\rho \mathscr{R}.
\end{equation}

Next, using the identity (\ref{iwe}) and fixing the gauge at $S_0=1$, we have 
\begin{equation}
  \bar{\mathscr{L}} = -[b- S -\bar{S}]_D + \left([a - \frac{1}{2 \rho} S^2]_F + \textrm{h.c.}\right)
\label{qet}
\end{equation}
and the corresponding K\"ahler potential and superpotential are
\begin{equation}
 K = -3 \ln (b - A - \bar{A})\quad,\quad  W = a - \frac{1}{2 \rho} A^2\, .
\end{equation}
The scalar potential $V$ is given by
\begin{align}
V = e^K \left[ g^{A \bar{A}}  (D_{A} W) (\bar{D}_{\bar{A}} \bar{W})  -3 \bar{W} W    \right]. 
\end{align}
Note that positivity of the kinetic terms implies that $b-2A_R > 0 $, where we have set $A = A_R + \mi A_I$. We now compute
\begin{equation}
\begin{gathered}
 e^K = \frac{1}{(b-A-\bar{A})^3} \textrm{ , }  g_{A \bar{A}} = \frac{\partial}{\partial A} \frac{\partial}{\partial \bar{A}} K = \frac{3}{(b- A-\bar{A})^2} \textrm{ , }  g^{A \bar{A}} =   \frac{(b - A - \bar{A})^2}{3}\, ,
\end{gathered}
\end{equation}
and
\begin{align}
 D_{A}W & = \partial_{A} W + K_{A} W  = - \frac{A}{\rho} + \frac{3}{b- A- \bar{A}} (a - \frac{1}{2 \rho} A^2). 
\end{align}
Putting everything together, we get that
\begin{align}
 V  & = \frac{A \bar{A}}{3 \rho^2 (b - A - \bar{A})} - \frac{A}{\rho (b - A -\bar{A})^2} (a - \frac{1}{2 \rho}\bar{A}^2) - \frac{\bar{A}}{\rho (b - A - \bar{A})^2} (a - \frac{1}{2\rho } A^2) \nonumber \\   
& = \frac{1}{\rho^2(b -2 A_R)^2} \left\{ \frac{1}{3} (A_R^2 + A_I^2) (b + A_R) - 2  a \rho A_R     \right\}\, .
\end{align}
The range of $A_R$ is given by
\begin{equation}
 - b \leqslant A_R < \frac{b}{2}\quad,\quad b>0\, ,
\label{dcon}
\end{equation}
so that the scalar potential is bounded from below.

To find the minimum of the potential, we demand that
\begin{equation}
\langle \frac{\partial V}{\partial A_R} \rangle = \langle \frac{\partial V}{\partial A_I} \rangle = 0.
\label{condas}
\end{equation}
The second of the equations above gives
\begin{equation}
   \langle   A_I (b+A_R)  \rangle = 0.
\label{eqq}
\end{equation}
\begin{table}
\centering
  \begin{tabular}{ c || c }
    $ \langle A_R \rangle $ &  $- b \leqslant A_R < \frac{b}{2}, b>0$ \\ \hline
     $ - \frac{b}{2}$ & true always \\ \hline
     $ b + \sqrt{b^2 - 6 a \rho}$, $b^2 - 6 a \rho \ge 0$ & never true  \\ \hline
     $ b - \sqrt{b^2 - 6 a \rho}$, $b^2 - 6 a \rho \ge 0$  & true if $b^2 > 8 a \rho$ and $b^2 \ge - 2 a \rho$  \\ 
  \end{tabular}
\caption[Table caption text]{Possible values of $\langle A_R \rangle$ for $\langle A_I \rangle = 0$.}
\label{aer2}
\end{table}
If $\langle A_R \rangle = -b$, then
\begin{equation}
\langle \frac{\partial V}{\partial A_R} \rangle = 0 \Rightarrow \langle A_I^2 \rangle = - 2 a \rho- b^2  \overset{\rho \rightarrow \infty}{\longrightarrow} - \infty, 
\end{equation}
 so this case is rejected. 
Consequently $\langle A_I \rangle = 0$. Then
\begin{equation}
 \langle \frac{\partial V}{\partial A_R} \rangle = 0 \Rightarrow \langle  (b+ 2 A_R) (A_R^2 - 2 b A_R + 6 a \rho ) \rangle = 0,
\end{equation}
which yields three solutions whose compatibility with the condition (\ref{dcon}) is given in Table~\ref{aer2}. Only the solutions $\langle A_R \rangle =  - \frac{b}{2}$ and $\langle A_R \rangle = b - \sqrt{b^2 - 6 a \rho}$ are compatible with the range of $A_R$. Now we would like to check whether one of them is compatible with the condition
\begin{equation}
\langle V \rangle = 0.
\label{cac}
\end{equation}
Equation (\ref{cac}) has two solutions whose compatibility with the condition (\ref{dcon}) is given in Table~\ref{aer3}. 
\begin{table}
\centering
  \begin{tabular}{ c || c }
    $ \langle A_R \rangle $ &  $- b \leqslant A_R < \frac{b}{2}, b>0$ \\ \hline
        $0$ & true always \\ \hline 
         $ -\frac{b}{2} +\frac{\sqrt{b^2 +24 a \rho}}{2}$, $b^2 + 24 a \rho \ge 0$ &  true if $b^2 > 8 a \rho$  \\ \hline 
     $  -\frac{b}{2} - \frac{\sqrt{b^2 +24 a \rho}}{2}$, $b^2 + 24 a \rho \ge 0$  & true if $a \rho \le 0$   \\ 
  \end{tabular}
\caption[Table caption text]{Possible values of $\langle A_R \rangle$ for $\langle V \rangle = 0$ and $\langle A_I \rangle = 0$.} 
\label{aer3}
\end{table}

It is straightforward to see that the solution $\langle A_R \rangle = b - \sqrt{b^2 - 6 a \rho}$ is compatible with (\ref{cac}) only if $b^2 = 8 a \rho$ (for $\langle A_R \rangle \neq 0$) or if $a=0$ (for $\langle A_R \rangle = 0$). The first case is rejected, since then $\langle A_R\rangle=b/2$ and the metric $g_{A \bar{A}}$ diverges. The second case is rejected, because then $\langle  D_{A} W \rangle = 0$ and there is thus no spontaneous supersymmetry breaking. 
On the other hand, the solution $ \langle A_R \rangle=-\frac{b}{2} $  is compatible with (\ref{cac}) for $b^2 +24 a \rho = 0$. It can also lead to spontaneous supersymmetry breaking, as
\begin{equation}
\langle e^{K/2} \sqrt{g^{A \bar{A}}} \bar{D}_{\bar{A}} \bar{W} \rangle \sim a b^{-3/2} \neq 0 \textrm{ for finite } \rho \, .   
\end{equation}
\begin{table}
\centering
  \begin{tabular}{ c || c || c }
    $g^{A \bar A} \frac{\partial^2 V}{\partial \phi_i \partial \phi_j} $ & $ A_R $ & $A_I$ \\ \hline
    $A_R$ & $- \frac{1}{\rho^2} \frac{b}{9}$ & $0$ \\ \hline 
    $A_I$ & $0$ & $\frac{1}{\rho^2} \frac{b}{9}$ \\
  \end{tabular}
\caption[Table caption text]{The (canonically normalized) scalar squared-mass matrix for $\langle A_R \rangle = -\frac{b}{2}$, $\langle A_I \rangle =0$ and $\langle V \rangle = 0$.}
\label{aer}
\end{table}
However, it is easy to see that the scalar squared-masses corresponding to $A_R$ and $A_I$ have opposite signs and thus the point ($\langle A_R \rangle = -\frac{b}{2}$, $\langle A_I \rangle =0$) is a saddle point of the potential and not a minimum, see Table~\ref{aer}. Moreover, all the eigenvalues of the scalar mass matrix approach $0$ as $|\rho| \rightarrow \infty$ and thus the extra scalar (sgoldstino) does not decouple. We conclude that neither of the two solutions for $\langle A_R \rangle$ can be used to tune the cosmological constant to zero for every value of $\rho$, consistently with the decoupling of the extra scalar.

Instead, we can investigate what happens if the condition (\ref{cac}) holds for the potential only in the limit $\rho \rightarrow \infty$. 
For both possible solutions 
\begin{equation}
\begin{gathered}
\langle A_I \rangle = 0 \quad,\quad  \langle A_R \rangle = b - \sqrt{b^2 - 6 a \rho} \approx \rho \lambda -1 + 3 \frac{W_0}{\lambda} \, , \lambda \neq 0  \\ 
\langle A_I \rangle = 0\quad,\quad \langle A_R \rangle = -\frac{b}{2} = - \frac{1}{2} - \rho \lambda
\label{solg}
\end{gathered}
\end{equation}
we find that $V \rightarrow 0$ for $ \rho \rightarrow \infty$; however, none of the eigenvalues of the scalar mass matrix approaches $\infty$ at $\rho \rightarrow \infty$ (they approach $0$ instead), which is again incompatible with the sgoldstino decoupling. We conclude that the parameter space of the model (\ref{baro}) does not allow for the realization of the non-linear supersymmetry coupled to gravity. Thus, (\ref{baro}) has to be modified suitably which is what we proposed in Section~\ref{tria}.


\newpage


\begin{thebibliography}{99}

\bibitem{Antoniadis:2010hs}
  I.~Antoniadis, E.~Dudas, D.~M.~Ghilencea and P.~Tziveloglou,
  \textit{Non-linear MSSM,}
  Nucl.\ Phys.\  B {\bf 841} (2010) 157
  [arXiv:1006.1662 [hep-ph]].

\bibitem{Antoniadis:2014oya}
  I.~Antoniadis, E.~Dudas, S.~Ferrara and A.~Sagnotti,
   \textit{The Volkov-Akulov-Starobinsky supergravity,}
  Phys.\ Lett.\ B {\bf 733} (2014) 32
  [arXiv:1403.3269 [hep-th]].

\bibitem{Ferrara:2014kva}
  S.~Ferrara, R.~Kallosh and A.~Linde,
   \textit{Cosmology with Nilpotent Superfields,}
  JHEP {\bf 1410} (2014) 143
  [arXiv:1408.4096 [hep-th]].
  
\bibitem{Bergshoeff:2015tra}
  E.~A.~Bergshoeff, D.~Z.~Freedman, R.~Kallosh and A.~Van Proeyen,
   \textit{Pure de Sitter Supergravity,}
  [arXiv:1507.08264 [hep-th]].

\bibitem{Antoniadis:1999xk}
  I.~Antoniadis, E.~Dudas and A.~Sagnotti,
   \textit{Brane supersymmetry breaking,}
  Phys.\ Lett.\ B {\bf 464} (1999) 38
  [hep-th/9908023].

\bibitem{Antoniadis:2004uk}
  I.~Antoniadis and M.~Tuckmantel,
   \textit{Nonlinear supersymmetry and intersecting D-branes,}
  Nucl.\ Phys.\ B {\bf 697} (2004) 3
  [hep-th/0406010].

  \bibitem{Kallosh:2015nia}
  R.~Kallosh, F.~Quevedo and A.~M.~Uranga,
   \textit{String Theory Realizations of the Nilpotent Goldstino,}
  [arXiv:1507.07556 [hep-th]].

\bibitem{Rocek:1978nb}
  M.~Rocek,
   \textit{Linearizing The Volkov-Akulov Model,}
  Phys.\ Rev.\ Lett.\  {\bf 41} (1978) 451.

\bibitem{Lindstrom:1979kq}
  U.~Lindstrom, M.~Rocek,
   \textit{Constrained Local Superfields,}
  Phys.\ Rev.\  D {\bf 19} (1979) 2300.

\bibitem{Casalbuoni:1988xh}
  R.~Casalbuoni, S.~De Curtis, D.~Dominici, F.~Feruglio and R.~Gatto,
  \textit{Nonlinear realization of supersymmetry algebra from supersymmetric constraint,}
  Phys.\ Lett.\  B {\bf 220} (1989) 569.

\bibitem{komar}
Z.~Komargodski and N.~Seiberg,
\textit{From Linear SUSY to Constrained Superfields,}
JHEP {\bf 0909} (2009) 066 [arXiv:0907.2441 [hep-th]].

\bibitem{kuzenko}
S. M.~Kuzenko and S. J.~Tyler, 
\textit{On the Goldstino actions and their symmetries,}
JHEP {\bf 1105},  (2011) 055 [arXiv:1102.3043 [hep-th]].

\bibitem{Volkov:1973ix}
  D.~V.~Volkov and V.~P.~Akulov,
 \textit{Is the Neutrino a Goldstone Particle?,}
  Phys.\ Lett.\  B {\bf 46} (1973) 109.

\bibitem{Antoniadis:2011xi}
  I.~Antoniadis, E.~Dudas and D.~M.~Ghilencea,
  \textit{Goldstino and sgoldstino in microscopic models and the constrained superfields formalism,}
  Nucl.\ Phys.\ B {\bf 857} (2012) 65
  [arXiv:1110.5939 [hep-th]].

\bibitem{Markou}
C.~Markou, Master's thesis, June 2015.

\bibitem{Dudas:2015eha}
  E.~Dudas, S.~Ferrara, A.~Kehagias and A.~Sagnotti,
  \textit{Properties of Nilpotent Supergravity,}
  [arXiv:1507.07842 [hep-th]].

\bibitem{Gates:2009hu}
  S.~J.~Gates, Jr. and S.~V.~Ketov,
  \textit{Superstring-inspired supergravity as the universal source of inflation and quintessence,}
  Phys.\ Lett.\ B {\bf 674} (2009) 59
  [arXiv:0901.2467 [hep-th]].

\bibitem{Ketov:2009wc}
  S.~V.~Ketov,
  \textit{Scalar potential in F(R) supergravity,}
  Class.\ Quant.\ Grav.\  {\bf 26} (2009) 135006
  [arXiv:0903.0251 [hep-th]].


\bibitem{super}
  R.~Kallosh, L.~Kofman, A.~D.~Linde and A.~Van Proeyen,
  \textit{Superconformal symmetry, supergravity and cosmology,}
  Class.\ Quant.\ Grav.\  {\bf 17} (2000) 4269
   [Class.\ Quant.\ Grav.\  {\bf 21} (2004) 5017]
  [hep-th/0006179].

\bibitem{freed}
D.Z. Freedman and A. Van Proeyen, \textit{Supergravity,} Cambridge University Press, Cambridge, UK, 2012.

\bibitem{cec}
  S.~Cecotti,
  \textit{Higher Derivative Supergravity Is Equivalent To Standard Supergravity Coupled To Matter. 1.,}
  Phys.\ Lett.\ B {\bf 190} (1987) 86.

\bibitem{ferr}
  S.~Ferrara, R.~Kallosh and A.~Van Proeyen,
  \textit{On the Supersymmetric Completion of $R+R^2$ Gravity and Cosmology,}
  JHEP {\bf 1311} (2013) 134
  [arXiv:1309.4052 [hep-th]].

\bibitem{wesbag}
J. Wess, J. Bagger, \textit{Supersymmetry and Supergravity,} Princeton University Press, Princeton, New Jersey, 1992.



\end{thebibliography}
\end{document}